\documentclass[twocolumn,superscriptaddress,groupedaddress]{revtex4}  
\usepackage{bookmark}
\usepackage{graphicx}  
\usepackage{dcolumn}   
\usepackage{bm}        
\usepackage{amssymb}   
\usepackage{amsmath}
\usepackage{feynmp}    
\begin{document}
\newcommand{\bq}{\begin{equation}}
\newcommand{\eq}{\end{equation}}
\newcommand{\bqn}{\begin{eqnarray}}
\newcommand{\eqn}{\end{eqnarray}}
\newcommand{\nb}{\nonumber}
\newcommand{\lb}{\label}

\title{Degenerate Dark Matter at Galactic Scales: A BCS Theory}
\author{Ahmad Borzou} \affiliation{EUCOS-CASPER, Physics Department, Baylor University, Waco, TX 76798-7316, USA}\email{ahmad\_borzou@baylor.edu}
\date{\today}

\begin{abstract}
We show that if dark matter in a typical galaxy is a degenerate Fermi gas, particles should have a mass of $\sim$ 1eV for the galaxy to be stable. 
While this is the mass range of the active neutrinos, they are not a dark matter candidate in SM-GR-$\Lambda$-CDM. To show that the bounds on active neutrino dark matter are model dependent, we explore the predictions of SM-LGT cosmological model.
First, primordial neutrinos are predicted to freeze-out non-relativistically at early universe without affecting the expansion rate.
Second, they make a degenerate gas in galaxies at the present time. 
Third, SM-LGT Hamiltonian at low temperatures is identical with that of the BCS theory of superconductivity. 
Consequently, there exists a narrow band of compressible condensed bosonic bound states on top of the Fermi surface which forms some denser structures. 
\end{abstract}

\maketitle

\section{Introduction}
A variety of independent observations leave no doubt on the existence of dark matter (DM). These include the early measurements of galaxies' velocity dispersion in the Coma cluster \cite{Zwicky1933}, the rotation curves in galaxies \cite{Rubin1980}, the recent measurements of the gravitational lensing \cite{Refregier2003,Tyson1998}, the Bullet cluster \cite{Clowe2006}, the anisotropies in the CMB \cite{Komatsu2011}, and the large scale structures \cite{Allen2003}. Cold non-interacting dark matter (CDM) fits well to data from extremely large structures down to distances of $\sim 10$ kpc. Nevertheless, below this scale there are several challenges \cite{Bullock2014} that call for more detailed model of dark matter. First, observations of the rotation curves indicate a core density \cite{Salucci2000,deBlok2008} which contradicts the cusp density predicted by CDM model \cite{Flores1994,Frenk1996b,Moore1999a}. Second, the number of observed sub-halos in the Milky Way are far less than what is predicted in a CDM scenario \cite{Moore1999b,Klypin1999}. Third, the super massive subhalos predicted by CDM have failed to attract visible matter \cite{Boylan2012,Parry2012}

Different solutions are suggested for the small scale problems of CDM. Visible matter feedback \cite{Eke1996,Governato2012}, warm dark matter \cite{Frenk2012}, and self-interacting dark matter \cite{Elbert2015,Tulin2018} are the popular ones. However, none is satisfactory enough. The feedback scenario works well for some galaxies but the problems persist in cases with little visible matter \cite{Ferrero2012}. The other two scenarios also---at least in their simple versions---can not explain the small scales without affecting the larger structures \cite{Pontzen2014}. 
A degenerate DM is another proposal that is put forward recently \cite{Paolo2018}. If DM is a degenerate Fermi gas in galaxies, the mass density tend to be lower in the inner regions while the cosmological predictions are not changed due to the absence of the degeneracy. 
In this paper we discuss a cosmological model that predicts a degenerate DM gas at galactic scales.

In addition to the small scale problems, CDM scenario confronts another challenge at this time. No candidate particle is observed in any of the experimental searches \cite{XENON2017,LUX2017,PandaX2017,Kahlhoefer2017,CMS2013,ATLAS2012}. The only stable and invisible particles that have been observed are active neutrinos. 
Due to several bounds, they are not however a DM candidate in SM-GR-$\Lambda$-CDM cosmological model---based on the standard model (SM), general relativity (GR), cosmological constant ($\Lambda$) and CDM. 
Nonetheless, the bounds are model dependent and can be removed in other cosmological models. One such model will be discussed in the following.

In this paper, we first show that if a typical galaxy is made of degenerate Fermi gas, it is stable only if DM has a mass of $\sim$1 eV, consistent with the mass of neutrinos. 
Next, after a review of Lorentz gauge theory of gravity (LGT) \cite{BorzouMain}, we study the galactic scale predictions of SM-LGT cosmological model \cite{BorzouCosmology1,BorzouCosmology2}. The first prediction is that primordial neutrinos are Fermi degenerate in galaxies. Second, primordial neutrinos become cold before freeze-out at the early universe without affecting the expansion of the universe. Therefore, neutrino dark matter scenario does not contradict the observations of structure formation in the universe \cite{Frenk1983,Dekel1984,Kaiser1983,Smoot1992,Planck2015}. 
Third, SM-LGT Hamiltonian at low temperatures is identical with the Hamiltonian of the BCS theory which together with the Fermi surface predict the existence of a narrow bosonic band---the same as what is observed in superconductors---that due to the absence of the degeneracy pressure can compress further and make the compact sub-halos that are often used to derive the phase-space bounds \cite{Tremaine,Hogan2000,Madsen2001,Dalcanton2001} on the mass of DM. The bounds however rely on the assumption that the sub-halos are made of fermions which is not valid here. This can provide a solution to the missing satellite problem since the number of dense satellite galaxies is a function of the width of the bosonic band and is limited. 

Finally, the vacuum energy does not gravitate in SM-LGT \cite{BorzouMain} alleviating the cosmological constant problem. 
Transition from decelerating to accelerating universe is spontaneous under the late universe conditions and there is no need for dark energy \cite{BorzouCosmology1}. 
While LGT has only one free parameter, it behaves much better than GR at the UV energy scales \cite{BorzouMain}.


\begin{figure}[b]
\includegraphics[scale=0.55]{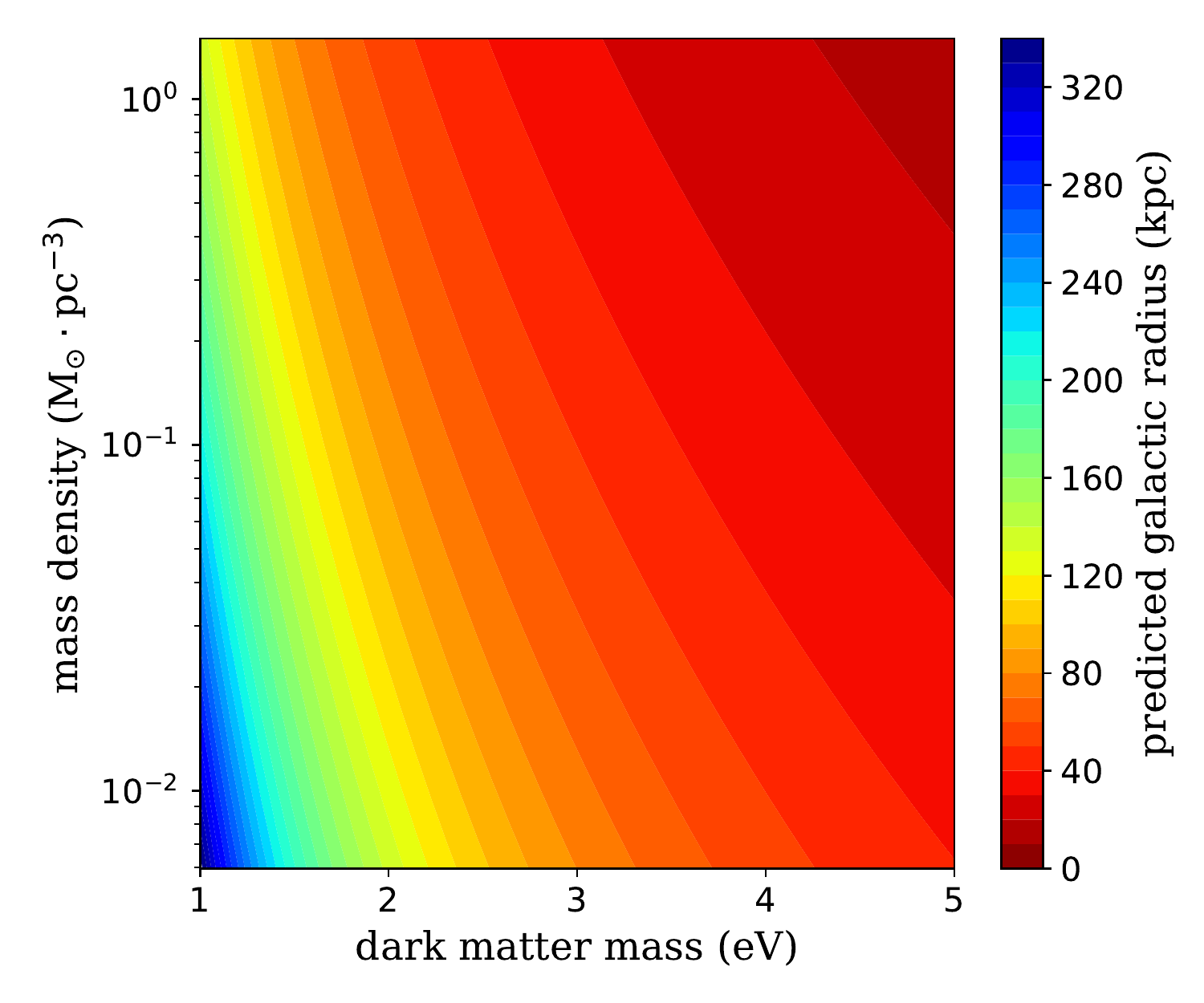}
\caption{\lb{Fig:m_rho_R}Predicted mass of dark matter in terms of the observed radius and mass density of a typical galaxy assuming that it is filled with degenerate Fermi gas.}
\end{figure}
\section{Stability condition}
Assuming that DM in galaxies is Fermi degenerate, their stability can be explored in a way similar to the study of white dwarfs \cite{Koester1990} by minimizing the free energy of the system which is the same as the total energy of the galaxy at zero temperature. 
Therefore, if the radius of the galaxy is altered infinitesimally, the change in the gravitational potential should be opposite of the work done by the pressure
\bqn
\lb{Eq:equilibrium}
\alpha G M^2/R^2 = 4\pi R^2 P,
\eqn
where $P$, $M$ and $R$ are the pressure, the mass and the radius of the galaxy respectively, $G$ is Newton's gravitational constant and $\alpha$ is a correction for inhomogeneities in the density distribution. The pressure of a degenerate Fermi gas is given by 
\bqn
&&P    = \frac{\pi m^4 c^5}{3h^3}A(x),
\eqn
where $A(x) = (2x^3-3x)\sqrt{x^2+1}+3\sinh^{-1} x$, and $x = \sqrt{\frac{2kT_f}{mc^2}}$. Here $m$ is the mass of DM, $c$ is the speed of light, $h$ is the Planck constant, $k$ is the Boltzmann constant, and $T_f$ is the Fermi temperature.
Inserting the pressure into the equilibrium equation \eqref{Eq:equilibrium}, the radius of the galaxy reads
\bqn
\lb{Eq:Galaxy_R}
R = \sqrt{\frac{3A(x)}{\alpha G}}\frac{m^2c^{\frac{5}{2}}}{2\rho h^{\frac{3}{2}}}.
\eqn
Since the density is known from observations and $A(x)$ is known for a given $m$, equation \ref{Eq:Galaxy_R} can be used to predict the mass of DM in terms of the radius of the galaxy which is plotted in Fig.~\ref{Fig:m_rho_R}. 
A galaxy like the Milky Way has a density of $10^{-2} M_\odot \cdot \text{pc}^{-3}$ and a radius of $\sim 100$ kpc predicting the dark matter mass to be in the range of $\sim 1$ eV consistent with the mass of active neutrinos. 
This is very surprising in the sense that a combination of the Planck constant ($10^{-33}\text{kg}\cdot\text{m}^2\cdot\text{s}^{-1}$), the speed of light ($10^{8}\text{m}\cdot\text{s}^{-1}$), the mass ($10^{42}\text{kg}$) and the density ($10^{-22}\text{kg}\cdot\text{m}^{-3}$) of a typical galaxy, and Newton's gravitational constant ($10^{-10}\text{m}^3\cdot\text{kg}^{-1}\cdot\text{s}^{-2}$) each raised to a different power return a value for the mass of dark matter ($10^{-36}\text{kg}$) which is roughly the mass of the only particles---active neutrinos---in SM that have the rest of the conditions for being dark matter, i.e. they do not interact with light and are stable. 
It becomes even more surprising when we note that the plot indicates a sensitive dependence of the radius of galaxy on the mass of DM.

\section{Lorentz gauge theory of gravity}
LGT is a Yang-Mills theory of gravity \cite{BorzouMain} with the homogeneous Lorentz as the gauge group and has a kinematics identical with general relativity which is described in \cite{Weinberg1972}. 
It is invariant under any local coordinate transformation and any local Lorentz transformation in the spinor space---the tangent spaces. The two spaces are related by the tetrad postulate. In GR the coordinate space is dynamical and the other space is determined by the tetrad postulate while in LGT the tangent space is dynamical and the space-time is determined by solving the tetrad postulate. This equivalently means that in GR energy-momentum tensor---corresponding with the coordinate invariance of the action---is the source of gravity. In LGT gravity is generated by the Lorentz current---corresponding with the Lorentz invariance in the spinor sector. 
Unlike the Poincare gauge theories, in LGT both the mass-generated and the spin-generated fields propagate while only two derivatives are involved on its dynamical field in its action 
\bqn
\lb{Eq:LGTAction}
S = \int e d^4x \left(-\frac{1}{4}F_{\alpha\beta mn}F^{\alpha\beta mn} + {\cal{L}}_{\text{SM}} + {\cal{L}}_{\text{Constraint}}\right),\nb\\
\eqn
where the last term is a Lagrange multiplier times the tetrad postulate---the covariant derivative of the tetrad $D_{\alpha}e_{m\beta}=0$.
The second term is the Lagrangian of the standard model. Finally $F_{\alpha\beta mn} = \partial_{\beta} A_{mn\alpha} - \partial_{\alpha} A_{mn\beta}+ gA_{m~\alpha}^{~~k} A_{kn\beta} - gA_{m~\beta}^{~~k} A_{kn\alpha}$ is the strength tensor defined by the commutation of two covariant derivatives and is equivalent with the Riemann curvature tensor if multiplied by two tetrads. Also, $A_{mn\alpha}$ is the gauge field in the spinor space. In the absence of matter---the last two terms in the action---LGT is identical with the Yang-Mills theories in the standard model \cite{BorzouPathIntegral} and therefore is renormalizable and unitary to all orders of perturbation. It should be noted that general relativity---or any of its alternatives that assume the energy-momentum tensor as the source of gravity---is not both unitary and renormalizable even in the absence of matter \cite{Hooft1974,Deser2000}. A notable difference between LGT and GR is that in the former the gauge field $A_{mn\alpha}$ is the dynamical field just like any Yang-Mills theory. In the latter however the metric which itself is a tensor under the Poincare symmetry of the space-time is the dynamical variable. This causes several problems. First, the vacuum energy is coupled with the determinant of the metric which is dynamical in GR. Therefore, it is expected to gravitate and leads to the cosmological constant problem \cite{Martin2012}. In LGT however the Lorentz gauge field---different than the Christoffel fields as the space-time gauge filed---is dynamical and the vacuum energy does not gravitate \cite{BorzouCosmology1}. Moreover, when the metric is dynamical the space-time is fundamentally discrete. This contradicts the basics of quantum mechanics with a continuous time \cite{Anderson2012}. 

When energies are well below the Planck energy scale, LGT field equations can be written as \cite{BorzouEffectiveLGT}
\bqn
\lb{Eq:FieldEq}
D^{\beta}F_{\alpha\beta mn} = -\frac{\delta {\cal{L}}_{\text{SM}}}{\delta A^{mn\alpha}} 
+ 4\pi G \left( D_n T_{\alpha m} - D_m T_{\alpha n} \right), 
\eqn
where $T$ is the fermionic part of the energy-momentum tensor---the bosonic part falls out for the same reason that the vacuum energy does---$G$ is Newton's constant and is an effective coupling derived after integrating over a Planckian length and can be expressed as the coupling constant of LGT $g$ multiplied by the Planckian length square. It is similar to the Fermi constant that is expressed as the coupling constant of the electroweak theory divided by the mass of the W boson which is integrated out of the underlying theory. For this reason the Fermi constant is extremely smaller than that of the electroweak. For the same reason $G \ll g$ in LGT.
The two terms on the right hand side can not be mixed under any of the transformations as the first is fully anti-symmetric---see equation \eqref{Eq:LGTInteraction}---while the second is partially symmetric. Therefore, two independent gravitational fields are being generated in LGT---only at the Planck energy scale the two fields are unified, the spin-generated gravitational field by the first source term and the familiar Newtonian field by the second. 
In the linear form and when the net spin of matter is zero, LGT and GR predict identical fields \cite{BorzouEffectiveLGT}. In the general non-linear form LGT does not reproduce GR. However, it also possesses very important exact solutions like the Schwarzschild, and the Kerr metrics \cite{BorzouEffectiveLGT}. Therefore, LGT passes all the experiments that GR has passed so far.

Any cosmological model should have a de Sitter like solution $\dot{a} \propto a$---with $a$ being the scale factor and dot indicating the time derivative---at the present epoch and a solution like $\dot{a} \propto a^{-1}$ at the early times. In GR one needs to assume an extremely fine tuned value for the vacuum energy to produce the former solution. To have the latter expression---the so called radiation dominated universe---as a solution in GR, one needs to assume that neutrinos are hot. None of the assumptions are experimentally verified. 
In LGT neither the vacuum energy nor the radiation does gravitate no matter how large or how small are they \cite{BorzouCosmology1}. 
Nevertheless, $\dot{a} \propto a^{\pm 1}$ are both the exact vacuum solutions of LGT, i.e. one does not need to make an assumption regarding the vacuum energy or the radiation in the universe if he wants to explain the same observations \cite{BorzouCosmology1}. 
A numerical study shows that under the conditions of the current epoch, a decelerating universe in LGT spontaneously becomes accelerating right at the expected time \cite{BorzouCosmology1}---alleviating the need for manipulating the vacuum energy. 

Since in LGT the expansion of the universe is not sensitive to the temperature of neutrinos, one of the constraints on the cold active neutrino dark matter scenario is removed. Moreover, since neutrinos are extremely light, a long-range force that is negligible with respect to the electroweak can delay their freeze-out in the early times \cite{BorzouCosmology2}. As was discussed above, in LGT there exists a spin-generated force of gravity that is orders of magnitude stronger than its mass-generated Newtonian force. It is shown that if neutrinos are as heavy as 1 eV, the coupling of LGT $g$ can be up to six orders of magnitude smaller than that of the electroweak but still strong enough to delay their freeze-out until the temperature of the universe is 1 eV and they are cold \cite{BorzouCosmology2}. Therefore, an active neutrino dark matter will no longer contradict the observations regarding the large scale structures in the universe \cite{Frenk1983,Dekel1984,Kaiser1983,Smoot1992,Planck2015}.
In the following we will show that the same force can alleviate the phase-space bounds on the cold active neutrino dark matter scenario. 

For energies well below the Planck mass, the only quantum mechanically important interaction between gravity---the spin generated sector---and matter in action \eqref{Eq:LGTAction} corresponding with the first term on the righ hand side of equation \eqref{Eq:FieldEq} is \cite{BorzouCosmology2}
\bqn
\lb{Eq:LGTInteraction}
{\cal{L}}_I = \frac{ig}{4}\epsilon_{kmn\alpha}A^{mn\alpha}\bar{\psi}\gamma^k\gamma^5\psi,
\eqn
where the Greek and the Latin indices have the same meaning and both run from 0 to 3 since the curvature of space-time is neglected here. Also, $\epsilon_{kmn\alpha}$ is the Levi-Civita symbol. The only difference between this term and the interaction in QED is the presence of $\gamma^5$ since the Levi-Civita symbol multiplied by the Lorentz gauge field is just a vector like the vector potential in electrodynamics. For static configurations the force is the same as the Coulomb force with the electric charge $e$ replaced by $g/4$ and is repulsive if the spin of particles are in the same direction and attractive if they are in the opposite directions \cite{BorzouCosmology2}.

\begin{figure}[t]
\includegraphics[scale=0.55]{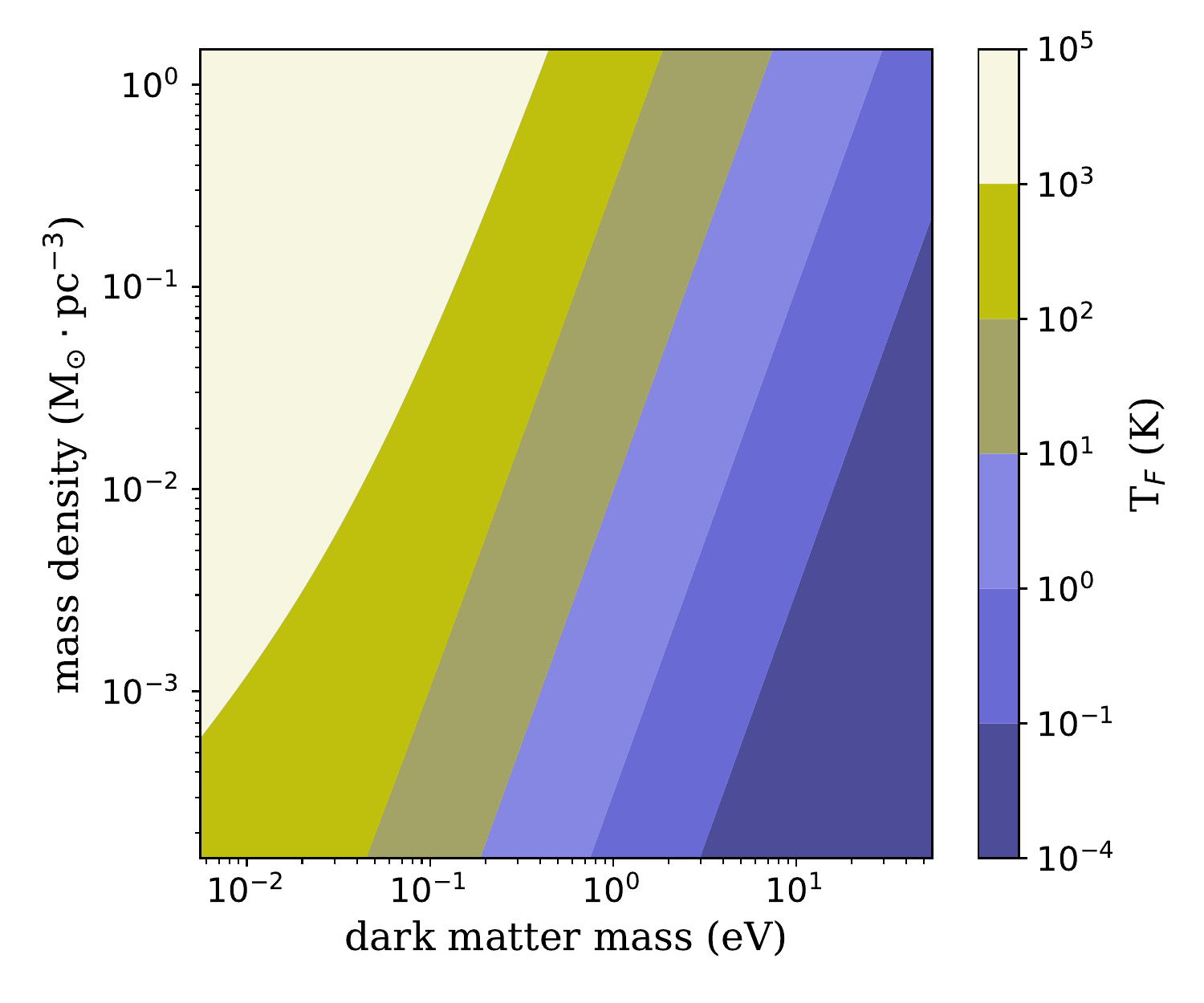}
\caption{\lb{Fig:FermiTemp}Fermi temperature in the color bar in terms of DM mass in the x-axis and density in the y-axis. If DM mass is $\sim$ 1 eV and temperature is $<10^{-4}$ Kelvins, DM halo is Fermi degenerate even for densities that are two orders of magnitude smaller than what is observed in the Milky Way $\sim 10^{-2} \text{M}_\odot\cdot\text{pc}^{-3}$.}
\end{figure}
\section{Temperature of neutrinos in galaxies}
In SM-GR-$\Lambda$-CDM, neutrinos freeze-out when the temperature is 1 MeV at the early universe with a relativistic distribution. Afterward they cool down due to the expansion of the universe. In this model their present temperature is predicted to be 1 MeV$\times a_{\text{freeze}}$ which is equivalent with 2 Kelvins. 

In SM-LGT neutrinos are predicted to freeze-out at 1 eV with a non-relativistic distribution. Their present temperature due to the expansion of the universe therefore should be 1 eV$\times a_{\text{freeze}}^2$ which in SI units is $\sim 10^{-4}$ Kelvins. 
The expansion of the universe is not the only way for neutrinos to cool themselves in SM-LGT. 
When neutrinos are trapped in the structures, they start to interact through the long-range spin-spin force in equation \eqref{Eq:LGTInteraction}. Therefore, they can cool down via radiation of the vector potential that was introduced in the same equation. This is similar to the Bremsstrahlung cooling that has been observed in electrically ionized galaxies. Since the power of this radiation is inversely proportional with the mass of neutrinos to power four and since neutrinos are extremely light, they should be even cooler than $10^{-4}$ Kelvins in galaxies.
This implies that neutrinos are very cooler than the visible matter in the galaxies and more importantly the observations regarding the kinetic motion of the visible matter in galaxies can not be directly extended to neutrinos as DM---something that is assumed in non-interacting models \cite{Ruchayskiy2009,Domcke2015}.

Neutrinos in galaxies are full Fermi degenerate if their temperature is well below the corresponding Fermi temperature---defined as the Fermi energy divided by the Boltzmann constant---which is model independent and a function of the density of the gas and the mass of the particles. In Fig.~\ref{Fig:FermiTemp} the Fermi temperature is shown by the colors for different DM masses in the x axis and typical mass densities in galaxies in the y axis. If neutrinos have a mass of 1 eV or lighter, their Fermi temperature is orders of magnitude larger than their predicted temperature of $< 10^{-4}$ Kelvins. 
The stability of such system was discussed above. Surprisingly, the system is stable only if the mass of DM is $\sim 1$ eV consistent with the prediction of SM-LGT.

\section{Statistics of neutrinos in galaxies}
\begin{figure}[b]
\includegraphics[scale=0.55]{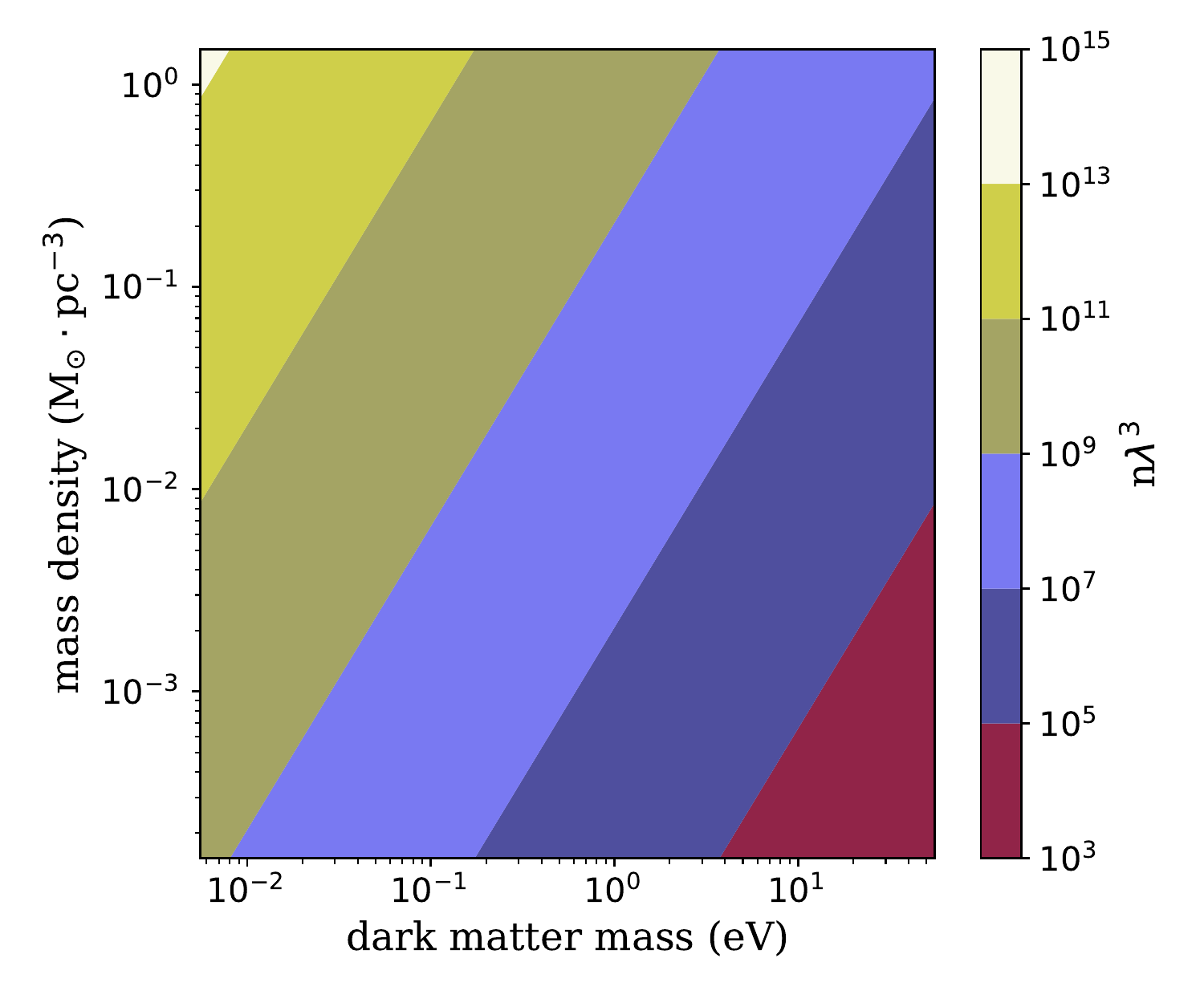}
\caption{\label{fig:nl3_m} Comparison of the thermal wavelength $\lambda$ at $10^{-4}$ Kelvins with the distance between particles n$^{-3}$ in terms of DM mass in the x-axis and density in the y-axis. Everywhere in the shown region the thermal wavelength is larger than the distance between particles and classical physics fail.}
\end{figure}
Whether a classical statistics is sufficient for describing the neutrino gas in galaxies or a full quantum statistics is needed can be decided by comparing the thermal wavelength $\lambda= h(2\pi m k T)^{-\frac{1}{2}}$---in terms of the Planck constant, the mass of neutrinos, the Boltzmann constant, and the temperature of the system---and the mean distance between the constituting particles $n^{-\frac{1}{3}}$. If the former is comparable to or larger than the latter, a full quantum statistics is needed. A ratio of the two raised to power three is shown in Fig.~\ref{fig:nl3_m}. For a particle in the mass range of neutrinos and densities that are typical in galaxies, a full quantum statistics shall be employed and any classical description of the neutrino gas---the dark matter halo in SM-LGT---fails.
Therefore, from here on we follow the method of quantized fields \cite{Fujita2002} where one does not need to assume a distribution for the neutrino gas as it will be derived from the basic principles and is the familiar Fermi-Dirac distribution only when the interactions are negligible. 

A straightforward calculation---see 
the appendix
for a detailed derivation---shows that at low temperatures the Hamiltonian corresponding with the Lagrangian in equation \eqref{Eq:LGTInteraction} is identical with that of the BCS theory of superconductivity \cite{Bardeen1957,Fujita2002}
\bqn
H = \sum_{k\sigma}\varepsilon_k c_k^{\dagger\sigma} c_k^{\sigma} - 
\frac{2U_0}{V}\sum_{k_1k_2}
c_{k_1}^{\dagger\uparrow}c_{-k_1}^{\dagger\downarrow}c_{-k_2}^{\downarrow}c_{k_2}^{\uparrow},
\eqn
where $c$ and $c^{\dagger}$ are the ladder operators of neutrinos, $U_0$ is the scattering length, $V$ is the volume, $\varepsilon_k$ is the energy of free particles, and $\sigma$ is the spin state. 

In superconductors that are described by the BCS theory, the attractive force among electrons is not sufficient to make a bound state. However, in the presence of the Fermi surface, the non-perturbative effects lead to creation of the bosonic pairon bound states regardless of the strength of the attractive force. Therefore, the electrons in superconductors are mostly free degenerate fermions accompanied with a narrow band of condensed bosonic electron-electron pairs with nearly zero net momentum on top of the Fermi surface. The pairons move through the crystal with no friction although their size is gigantic. 

Since the Hamiltonian of SM-LGT is the same as that of the BCS theory and since in SM-LGT also a Fermi surface is predicted in galaxies, the physics of neutrinos in galaxies is the same as that of electrons in superconductors. Majority of neutrinos are in the form of degenerate Fermi gas and as we showed above, generate a degeneracy pressure that supports the galaxy from collapsing under its own gravity. These neutrinos make the halo of the main galaxies and resist compression. On the other hand, there is also a narrow band of condensed bosonic neutrino-neutrino pairs in each galaxy on top of the Fermi surface which can compress further due to the absence of the degeneracy pressure. Therefore, these should be identified as the denser parts of or structures within a galaxy. 
This prediction may explain the missing satellite problem since the mass of the narrow bosonic band is limited and negligible with respect to the mass of the Fermi sea. Therefore, there is a limitation on the number of denser sub-structures in a galaxy.  
Also, the bosonic pairs can not compress more than a level. Because, if the pressure exceeds a threshold, the bound states are expected to break down leading to a degeneracy pressure that shoots the particles out of the core, providing a potential explanation for the core-cusp problem.

In the end we would like to discuss the phase-space lower-bounds on the mass of DM. If DM halo is a non-degenerate Fermi gas, the bounds are rather stringent \cite{Tremaine,Hogan2000,Madsen2001,Dalcanton2001}. They are however significantly lower for degenerate Fermi DM halos. In this case it is shown that the lower bound on the mass of DM is as low as a few tens of eV \cite{Paolo2018}. In SM-LGT this bound is even lower since the denser regions of galaxies that are used to derive the bound are made of condensed bosonic pairs.

\section{Conclusions}
We have studied the stability of a model independent Fermi degenerate dark matter halo and have shown that galaxies with the typically observed densities and radii are stable if DM mass is $\sim 1$ eV. Although the predicted DM mass is in the range of active neutrinos in SM, they are not a DM candidate in SM-GR-$\Lambda$-CDM cosmological model due to several contradictions with observations. 

In this paper we have shown that the bounds are model dependent and may not exist in other cosmological models. We have specially derived the predictions of SM-LGT cosmological model and have shown that primordial neutrinos freeze-out cold at early universe and are at a temperature below $10^{-4}$ Kelvins at the current epoch. Therefore, if the observed DM densities are all coming from neutrinos, they are Fermi degenerate in galaxies. Prediction of Fermi degeneracy of DM halos made of neutrinos is consistent with the DM mass that was derived in the model independent stability analysis. 

It also has been shown that at low temperatures the Hamiltonian of SM-LGT is the same as the Hamiltonian of the BCS theory. As a result, the physics of DM in galaxies should be very similar to the physics of superconductors. The most notable consequence of which is that the vacuum state of the theory changes near the Fermi surface indicating the existence of bosonic bound-states---although negligible in comparison with the number of fermions below the Fermi surface. Due to the absence of the degeneracy pressure the pairons can be compressed and make the denser regions or structures that are observed within galaxies. It also has been discussed that the model has potentials for solving the small scale problems of DM.

~\\{\bf Acknowledgements:}
We are grateful to M. Khodadadi Fard and M. Amini for bringing it to our attention that the Kerr metric is an exact solution to LGT.

\appendix
\section*{Appendix A: Hamiltonian of LGT at Low Temperatures}
The Hamiltonian corresponding with the Lagrangian in \eqref{Eq:LGTInteraction} is
\bqn
\text{H} = \int d^3x \left(\dot{\psi}\Pi - {\cal{L}}\right),
\eqn
where $\Pi = \partial{\cal{L}}/\partial\dot{\psi}=i\psi^{\dagger}$. This can be divided into a free and an interaction term $H = H_0 + H_{\text{I}}$ which are given by
\bqn
&&H_0 = \int d^3x \psi^{\dagger}\left( -i\gamma^0\sum_{\mu=1}^3\gamma^{\mu}\partial_{\mu} + m\gamma^0\right)\psi,\nb\\
&&H_{\text{I}} =\frac{g}{4}\epsilon^{k\mu ij} \int d^3x A_{ij\mu}\bar{\psi}\gamma_k\gamma^5\psi.
\eqn
If anti-particles are absent in the system, the fermioninc fields can be expanded as 
\bqn
\psi = \sum_{k=-\infty}^{\infty}\sum_{\sigma=1}^{2} \left(\frac{m}{V\varepsilon_k}\right)^{\frac{1}{2}}c_k^{\sigma}u^{\sigma}(k)e^{-i\vec{k}\cdot\vec{x}},
\eqn
where $V$ is the volume, $\varepsilon_k$ is the energy of free particles, $m$ is the mass, and $\sigma$ is the spin state. The anti-commutation of the ladder operators read $\{c_k^{\sigma},c_{k'}^{\dagger\sigma'}\} = \delta^{\sigma\sigma'}\delta_{kk'}$. A substitution of this into the free Hamiltonian and using $u^{\dagger\sigma}_ku^{\sigma'}_k= \frac{\varepsilon_k}{m}\delta^{\sigma\sigma'}$ reads
\bqn
H_0 = \sum_{k\sigma}\varepsilon_k c_k^{\dagger\sigma} c_k^{\sigma},
\eqn
which is a familiar term. To find the interaction Hamiltonian, we first need to find an expression for $A_{ij\mu}$. This can be found by solving the field equation \eqref{Eq:FieldEq} in the Lorentz gauge $\partial^{\mu}A_{ij\mu}=0$
\bqn
\partial^2A^{ij\mu} = \frac{g}{4} \epsilon^{k\mu ij}\bar{\psi}\gamma_k\gamma^5\psi.
\eqn
The solution can be expressed using a Green's function
\bqn
A^{ij\mu} = \frac{g}{4}\epsilon^{k\mu ij}\int d^3x' G\left(\vec{x}-\vec{x'}\right)\bar{\psi}(x')\gamma_k\gamma^5\psi(x').
\eqn
The interaction Hamiltonian now reads
\bqn
H_{\text{I}} &=& \left(\frac{g}{4}\right)^2 \sum_{k_1k_2k_3k_4}\sum_{\sigma_1\sigma_2\sigma_3\sigma_4} \frac{m^2}{V^2\sqrt{\varepsilon_{k_1}\varepsilon_{k_2}\varepsilon_{k_3}\varepsilon_{k_4}}}\nb\\
&&c_{k_1}^{\dagger\sigma_1}c_{k_2}^{\sigma_2}c_{k_3}^{\dagger\sigma_3}c_{k_4}^{\sigma_4}
\times\bar{u}^{\sigma_1}_{k_1}\gamma_k\gamma^5u^{\sigma_2}_{k_2}
\bar{u}^{\sigma_3}_{k_3}\gamma^k\gamma^5u^{\sigma_4}_{k_4}\nb\\
&&\int d^3x d^3x' G(\vec{x}-\vec{x'}) e^{i(\vec{k}_1-\vec{k}_2)\cdot(\vec{x}-\vec{x'})}.
\eqn
This expression can be simplified by defining $\vec{r}\equiv \vec{x}-\vec{x'}$ and $\vec{R}\equiv \frac{1}{2}\left(\vec{x}+\vec{x'}\right)$ and using $\int d^3R = V$. Moreover, at very low energies $\varepsilon_k \sim m$ and the leading term in $\bar{u}^{\sigma_1}_{k_1}\gamma_k\gamma^5u^{\sigma_2}_{k_2}
\bar{u}^{\sigma_3}_{k_3}\gamma^k\gamma^5u^{\sigma_4}_{k_4}$ reads 
$(-1)^{\sigma_1+\sigma_3}\delta^{\sigma_1\sigma_2}\delta^{\sigma_3\sigma_4}$. 
After all these simplifications the Hamiltonian takes the following form
\bqn
H_{\text{I}} = \frac{1}{V}\sum_{k_1k_2k_3k_4}\sum_{\sigma_1\sigma_3}(-1)^{\sigma_1+\sigma_3}
c_{k_1}^{\dagger\sigma_1}c_{k_2}^{\sigma_1}c_{k_3}^{\dagger\sigma_3}c_{k_4}^{\sigma_3}
~U\left(k_1,k_2\right),\nb\\
\eqn
where 
\bqn
U\left(k_1,k_2\right) \equiv 4\pi\left(\frac{g}{4}\right)^2\int rdr G(\vec{r})
\frac{\sin(|\vec{k}_1-\vec{k}_2|r)}{|\vec{k}_1-\vec{k}_2|r}.
\eqn
At very low energies $\sin(|\vec{k}_1-\vec{k}_2|r)\sim|\vec{k}_1-\vec{k}_2|r$
and the term becomes momentum independent
\bqn
U_0 = 4\pi\left(\frac{g}{4}\right)^2\int rdr G(\vec{r}). 
\eqn
Moreover, the net momentum before and after collisions should be zero, i.e. $\vec{k}_1+\vec{k}_3=\vec{k}_2+\vec{k}_4=0$. Inserting these two simplifications the Hamiltonian reads
\bqn
H_{\text{I}} = \frac{~U_0}{V}\sum_{k_1k_2}\sum_{\sigma_1\sigma_3}(-1)^{\sigma_1+\sigma_3}
c_{k_1}^{\dagger\sigma_1}c_{-k_1}^{\dagger\sigma_3}c_{-k_2}^{\sigma_3}c_{k_2}^{\sigma_1}.
\eqn
Since the sum is invariant under $k_2 \rightarrow -k_2$ and using the anti-commutation of the operators, it can be shown that the non-zero terms are those with opposite spin states. Also, the resulting two terms are equal and the Hamiltonian takes the following form
\bqn
H = \sum_{k\sigma}\varepsilon_k c_k^{\dagger\sigma} c_k^{\sigma} - 
\frac{2U_0}{V}\sum_{k_1k_2}
c_{k_1}^{\dagger\uparrow}c_{-k_1}^{\dagger\downarrow}c_{-k_2}^{\downarrow}c_{k_2}^{\uparrow}.
\eqn
This is identical with the Hamiltonian that is proposed for the BCS theory of superconductivity \cite{Bardeen1957,Fujita2002}. From here on we can simply follow the literature. The first step is to use the mean field approximation 
\bqn
AB \simeq \langle A\rangle B + A\langle B\rangle - \langle A \rangle \langle B \rangle,
\eqn
where $A$ and $B$ are two operators and $\langle \rangle$ indicates an ensemble average. Defining $A \equiv c_{k_1}^{\dagger\uparrow}c_{-k_1}^{\dagger\downarrow} $, $B \equiv c_{-k_2}^{\downarrow}c_{k_2}^{\uparrow}$, 
and $\Delta_k \equiv \frac{2U_0}{V}\sum_{k'}\langle c_{-k'}^{\downarrow}c_{k'}^{\uparrow} \rangle$, 
the Hamiltonian reads
\bqn
H &=& \sum_{k\sigma}\varepsilon_k c_k^{\dagger\sigma} c_k^{\sigma}
+ \frac{2U_0}{V}\sum_{kk'}
\langle c_{k}^{\dagger\uparrow}c_{-k}^{\dagger\downarrow}\rangle 
\langle c_{-k'}^{\downarrow}c_{k'}^{\uparrow}\rangle\nb\\
&-&\sum_{k} \Delta_k c_{k}^{\dagger\uparrow}c_{-k}^{\dagger\downarrow}
-\sum_{k} \Delta_k^{\ast} c_{-k}^{\downarrow}c_{k}^{\uparrow}.
\eqn
The second term is only a constant and does not change the dynamics of the system. 
The Hamiltonian is bilinear in the ladder operators and can be diagonalized through the Bogoliubov transformation
\bqn
\begin{pmatrix}
\gamma^{\uparrow}_k \\
~\\
\gamma^{\dagger\downarrow}_{-k}\\
\end{pmatrix}
\equiv
\begin{pmatrix}
u_k^{\ast} & -v_k \\
~ & ~ \\
v_k^{\ast} & u_k \\
\end{pmatrix}
\begin{pmatrix}
c^{\uparrow}_k \\
~\\
c^{\dagger\downarrow}_{-k}\\
\end{pmatrix},
\eqn
where $\{\gamma^{\sigma}_k,\gamma^{\dagger\sigma'}_k\}=\delta^{\sigma\sigma'}$. Also, $u_k$ and $v_k$ are chosen such that the Hamiltonian is diagonal
\bqn
&&|u_k|^2 =\frac{1}{2}\left(1+ \frac{\varepsilon_k}{\sqrt{\varepsilon_k^2 + \Delta_k^2}} \right)\nb\\
&&|v_k|^2 =\frac{1}{2}\left(1- \frac{\varepsilon_k}{\sqrt{\varepsilon_k^2 + \Delta_k^2}} \right)\nb\\
&&u_kv_k = \frac{\Delta_k}{2\sqrt{\varepsilon_k^2 + \Delta_k^2}}.
\eqn
Inserting all the pieces into the Hamiltonian, it takes its diagonal form
\bqn
H &=& 
\frac{2U_0}{V}\sum_{kk'}
\langle c_{k}^{\dagger\uparrow}c_{-k}^{\dagger\downarrow}\rangle 
\langle c_{-k'}^{\downarrow}c_{k'}^{\uparrow}\rangle +
\sum_{k\sigma}\sqrt{\varepsilon_k^2 + \Delta_k^2} \gamma^{\dagger\sigma} \gamma^{\sigma}.\nb\\
\eqn
Since $\gamma^{\dagger\sigma} \gamma^{\sigma}$ is the number operator for quasi-neutrinos, its ensemble expectation should be that of any fermion but with the new eigen-energies
$\left(\exp\left(\beta \sqrt{\varepsilon_k^2 + \Delta_k^2}\right)+1\right)^{-1}$ with $\beta$ being the inverse of the temperature multiplied by the Boltzmann constant. Also, the chemical potential is zero which is because the quasi-neutrinos are frequently created out of the pairon band and absorbed into the band. If the temperature is exactly zero, there should exist no quasi-neutrino at all. This distribution however can be used to calculate the ensemble expectation of the original ladder operators 
\bqn
\langle c^{\downarrow}_{-k}c^{\uparrow}_k\rangle = \frac{\Delta_k}{2\sqrt{\varepsilon_k^2+\Delta_k^2}}\left(1-\frac{2}{e^{\beta\sqrt{\varepsilon_k^2+\Delta_k^2}}+1}\right).
\eqn
Inserting this into the definition of $\Delta_k$, the so called gap equation now reads 
\bqn
\Delta_k = \frac{2U_0}{V}\sum_{k}
\frac{\Delta_k}{2\sqrt{\varepsilon_k^2+\Delta_k^2}}\left(1-\frac{2}{e^{\beta\sqrt{\varepsilon_k^2+\Delta_k^2}}+1}\right).
\eqn
The spin-spin interactions among neutrinos are weak enough to be neglected in perturbative regions. However, near the Fermi surface the effects are non-perturbative and can not be ignored. Therefore, we can assume that $U_0$ is non-zero only when $|\varepsilon_k - \varepsilon_{\text{F}}|<\omega_D$ where $\varepsilon_{\text{F}}$ is the Fermi energy and $\omega_D$ is the radial width of the bosonic band on top of the Fermi sea and is determined by the strength of the spin-spin interaction among neutrinos, i.e. the coupling constant of LGT $g$. 
If $\Delta_k = \Delta_0$ in the bosoninc band---it is clearly zero outside the band---and at zero temperature the equation reads
\bqn
1 = D(\varepsilon_{\text{F}})U_0\int_{\varepsilon_{\text{F}}-\omega_D}^{\varepsilon_{\text{F}}+\omega_D}  
\frac{d\varepsilon}{\sqrt{\varepsilon^2+\Delta_0^2}},
\eqn
where the density of states $D(\varepsilon)$ is approximated at the Fermi surface and taken out of the integral since $\omega_D \ll \varepsilon_{\text{F}}$. The non-perturbative effect can be seen by noting that the integral is in fact $\sinh^{-1}(\varepsilon/\omega_D)$.

\end{document}